\documentclass[12pt,preprint]{aastex}

\begin{document}

\title{Identifying Bright Stars in Crowded Environments Using Velocity Dispersion 
Measurements, and an Application to the Center of M32}

\author{T. J. Davidge }

\affil{Herzberg Institute of Astrophysics,
\\National Research Council of Canada, 5071 West Saanich Road,
\\Victoria, B.C. Canada V9E 2E7\\ {\it email: tim.davidge@nrc.ca}}

\author{Tracy L. Beck}

\affil{Space Telescope Science Institute,\\3700 San Martin Drive,\\
Baltimore, MD 21218 \\ {\it email: tbeck@stsci.edu}}

\author{Peter J. McGregor}

\affil{Research School of Astronomy and Astrophysics,
\\ Australian National University, Cotter Road, Weston Creek,
\\ ACT 2611, Australia \\ {\it email: peter@mso.anu.edu.au}}

\altaffiltext{1}{Based on observations obtained at the
Gemini Observatory, which is operated by the Association of Universities
for Research in Astronomy, Inc., under a co-operative agreement with the
NSF on behalf of the Gemini partnership: the National Science Foundation
(United States), the Science and Technology Facilities Council
(United Kingdom), the National Research Council of Canada (Canada), CONICYT (Chile), 
the Australian Research Council (Australia), the Ministerio da Ciencia e Technologia (Brazil),
and the Ministerio de Ciencia, Tecnologia e Innovacion Productiva (Argentina).}

\begin{abstract}

	The identification of individual stars in crowded environments using 
photometric information alone is confounded by source confusion. However, with the addition of 
spectroscopic information it is possible to distinguish between blends and areas where 
the light is dominated by a single star using the widths of absorption features. We describe a
procedure for identifying locations in kinematically hot environments where the light 
is dominated by a single star, and apply this method to spectra with 0.1 arcsec 
angular resolution covering the $2.1 - 2.3\mu$m interval in the central regions of M32. 
Targets for detailed investigation are selected as areas of localized brightness 
enhancement. Three locations where at least 60\% of the $K-$band light 
comes from a single bright star, and another with light 
that is dominated by two stars with very different velocities, are identified. The 
dominant stars are evolving near the tip of the asymptotic giant branch (AGB), and 
have M5 III spectral type. The lack of a dispersion in spectral-type suggests that
the upper AGB within the central arcsec of M32 has a dispersion in $J-K$ of 
only a few hundreths of a magnitude, in agreement with what is seen at larger 
radii. One star has weaker atomic absorption lines than the others, such 
that [M/H] is 0.2 dex lower. Such a difference in metallicity is 
consistent with the metallicity dispersion inferred from the width of the AGB in M32. The use of 
line width to distinguish between blends involving many relatively faint stars, none of which 
dominate the light output, and areas that are dominated by a single intrinsically bright 
star could be extended to crowded environments in other nearby galaxies.

\end{abstract}


\section{INTRODUCTION}

	Stellar content studies of nearby galaxies are of fundamental importance for 
furthering our understanding of galaxy evolution. Unfortunately, source confusion presents 
a formidable obstacle for efforts to identify individual stars in even the closest galaxies. 
Stars that fall within a single angular resolution element will appear as a single source, 
with spectrophotometric properties that are a luminosity-weighted amalgam of the component 
stars. With the possible exception of the outermost regions of galaxies, almost all 
stars observed in external galaxies are blends, although in most cases these 
involve an intrinsically bright star and a much fainter star (or stars), such that the 
combined light can be considered to be that of the brighter source. A more insidious 
situation occurs when sources with similar properties are blended. Consider the implications 
on a stellar content study if two stars with the same brightness and color 
are blended together. The blend will be brighter than the component stars but have the 
same color, and such an object may be mis-interpreted as evidence 
for a younger and/or more metal-poor population than that of the component 
stars. While the incidence of blending among stars with comparable 
brightness drops markedly for objects that are in the most advanced stages of evolution, 
because such objects are relatively rare, this remains an issue for studies of the central 
regions of even the nearest galaxies (Renzini 1998).

	There is a continuum in the severity of blending, and the key for stellar content 
studies is to identify locations in galaxies where the light output is dominated by one star.
In this paper we discuss a technique for identifying such objects in crowded, kinematically 
hot environments. The core principle is that the spectra of blends in 
which no single star dominates the light output will have line widths that are defined 
by the velocities of the component stars, and hence that are comparable to the 
local velocity dispersion in the galaxy. Spectra of blends that are dominated by one star 
will have line widths that are considerably narrower. The application of 
the technique is demonstrated using observations of  
the center of the Local Group compact elliptical (cE) galaxy M32. M32 is an 
unsurpassed laboratory for honing techniques that are used to explore the star-forming histories 
of galaxies. It is at a favorable distance such that light from a representative sample 
of stars passes through a spectrograph slit with a width of a few arcsec, while 
individual stars can also be resolved throughout much of the galaxy. As a result, 
the stellar content of M32 can be probed by investigating the 
properties of both resolved stars and integrated light. In addition,
the low interstellar gas content (e.g. Welch \& Sage 2001) suggests that M32 is free 
of the dust that can complicate stellar content studies at visible wavelengths. 
Finally, the brightest stars in M32 have $K \sim 15.5$ (Davidge 2000; Davidge et al. 
2000; Davidge \& Jensen 2007), and hence are well within the realm of being studied 
spectroscopically with existing telescopes. 

	Numerous studies have concluded that M32 is not a uniformly old stellar 
system, and contains a significant population of stars that formed within the past 
few Gyr, but not as recently as in the LMC. Indeed, the peak AGB brightness in M32 
is M$_K \approx -9$, which is $\approx 1$ magnitude fainter than the peak 
brightness seen among long period variables (LPVs) in the LMC (Fraser et al. 2005). 
Panoramic surveys have also found a number of AGB stars in the LMC that are {\it at least} as 
bright as those in M32 (Nikolaev \& Weinberg 2000; Cioni et al. 2006). With M$_K \approx -9$, 
the peak AGB brightness in M32 is comparable to that near the 
center of M31 (Davidge 2001), where there are spectroscopic signatures of an intermediate age 
population (e.g. Davidge 1997; Sil'Chenko, Burenkov, \& Vlasyuk 
1998). Adopting a distance modulus for the Galactic Center of 
14.5, then the two brightest AGB stars in the sample of Galactic Bulge stars studied by 
Frogel \& Whitford (1987), which are numbers 239 and 181 in their target list, 
have M$_K$ between --9 and --9.5. Many of the brightest AGB stars in M32 are 
LPVs with amplitudes of 1.0 magnitude in $K$ (Davidge \& Rigaut 2004), 
and so the non-variable AGB-tip in M32 corresponds to M$_K \approx -8.5$. Isochrones with 
solar metallicity from the compilation discussed by Girardi et al. (2002) indicate that this 
AGB-tip brightness is appropriate for a system with an age that is in excess of 1 Gyr, 
in agreement with studies of the integrated spectrum 
of M32 (O'Connell 1980; Davidge 1990; Bica, Alloin, \& Schmidt 
1990; del Burgo et al. 2001; Worthey 2004; Rose et al. 2005). 

	In the present study, three locations that are dominated by light from one star are 
identified in the central arcsec of M32, while a fourth location that is dominated by light 
from two stars that have very different velocities 
is also identified. These are the closest-in individual 
stars to be detected near the center of M32 at visible/near-infrared wavelengths. 
While five stars constitute an obviously limited sample, the data are sufficient 
to demonstrate the application of the technique. Even though the number of sources that 
are dominated by a single star and that can be identified at the diffraction limit of an 
8 metre telescope near the center of nearby galaxies like M32 is modest, a 20 -- 30 
metre telescope that delivers moderately high Strehl ratios should be able to detect 
roughly an order of magnitude more resolved sources. 

\section{OBSERVATIONS \& REDUCTIONS}

	The data used in this paper were discussed previously by Davidge et al. 
(2008) in their investigation of cE galaxies. 
In brief, spectra of M32 were recorded with NIFS$+$ALTAIR on Gemini North (GN) 
on the night of October 23, 2005 during NIFS commissioning. ALTAIR (Herriot et al. 
2000) is the facility AO system at GN, while NIFS (McGregor et al. 2003) is an 
integral field spectrograph for use in the $0.9 - 2.5\mu$m wavelength interval. NIFS samples 
a $3 \times 3$ arcsec$^2$ area on the sky with 29 $0.1 \times 3$ arcsec$^2$ slitlets. 
A spectrum with a dispersion 5300 is recorded on the $2048 \times 2048$ HAWAII-2RG HgCdTe 
detector in one of three ($J, H$, or $K$) atmospheric windows during a single exposure.

	The bright semi-stellar nucleus of M32 served as the reference source for 
AO correction. Six 600 sec exposures were recorded in the $K-$band, and 
each observation of M32 was followed by that of a blank sky field. 
Each M32 $+$ sky pair was differenced, and the results
were divided by a flat-field frame. The flat-fielded data were
wavelength calibrated using an Ar arc observation that was 
recorded on the same night as the M32 data. The wavelength-calibrated spectra
were summed and then divided by the spectrum of a telluric standard star, 
which was observed immediately following the M32 observations.

	There are a number of distinct sources with FWHM $\sim 0.1$ arcsec 
in the two-dimensional NIFS spectra, and these define the locations that are 
investigated in the remainder of the paper. The locations of these point 
sources are indicated in Figure 1, where (X,Y) = 
(0,0) is the center of M32, while the angular offsets of the sources from the center of M32 
are listed in Table 1. The detection of objects along the X = 0 axis is hindered by elevated 
star counts along the major axis of M32, which runs close to X = 
0 arcsec, and scattered light from the bright nucleus of M32.

	A spectrum was extracted of each of the sources in Figure 1. The surface brightness 
in the central arcsec of M32 exceeds 11 mag arcsec$^{-2}$ in $K$ (e.g. Jarrett et al. 
2003), and thus contributes substantial noise to the extracted spectra. Indeed, the 
unresolved stellar background, rather than the ambient `sky', is the dominant source of noise
in studies of stars near the centers of most nearby galaxies. A mean background 
spectrum was constructed from galaxy light bracketing each source, and the result was 
subtracted from the extracted spectrum. The background-subtracted spectra were 
binned along the dispersion axis to increase the signal-to-noise ratio, and 
the final spectral resolution is $\approx 1000$. The continuum was 
also removed from the spectra in preparation for the velocity dispersion analysis.

\section{VELOCITY MEASUREMENTS}

	With an image quality of FWHM $\approx 0.1$ arcsec, each angular resolution element 
near the center of M32 samples an area with an integrated magnitude M$_K \approx -9$, which is 
comparable to that of a Galactic globular cluster. Consequently, all of the 
sources in Figure 1 are blends, in the sense that light from more than one star falls within 
each angular resolution element. In addition, only objects that have an intrinsic brightness 
that is comparable to or greater than that in each angular resolution element 
(i.e. M$_K \approx -9$ in this case) will be resolved. The AGB-tip in M32 occurs near K = 15.5, 
which corresponds to M$_K \approx -9$ (Davidge et al. 2000; Davidge \& Jensen 
2007), and so any sources identified here are evolving near the AGB-tip. 

	Are there locations in the NIFS dataset where the light is dominated by a single bright 
star? Line width measurements are one way to determine if such locations are present. If the 
light in a blend originates from a large number of stars, with no single star contributing 
more than a small fraction of the light, then the absorption lines in the 
composite spectrum will be blurred by an amount that is comparable to the local velocity 
dispersion, $\sigma$, which is $\sim 100$ km sec$^{-1}$ near the edges of the NIFS field. 
At the other extreme, if the light comes from only one star then the absorption lines 
will be narrow, with the line width defined by the spectral resolution of the 
instrument. There are a number of techniques for measuring $\sigma$, and in this study 
the Tonry \& Davis (1979) routine, as implemented in IRAF, is employed. 

	A key issue is translating velocity dispersion into a measure of the amount of 
light that a bright single source contributes to the total light from a resolution 
element. Consider the spectrum of a location S$_{loc}$ on the sky that contains light from 
a star S$_{*}$, which contributes a fraction $f$ of the light, combined with that of a 
population of other fainter objects S$_{other}$, which together contribute a 
fraction $(1-f)$ of the light. The reader is reminded that S$_{loc}$, S$_{*}$,
S$_{other}$, and $f$ are all functions of wavelength, but this is not indicated here in 
the interest of brevity. The spectrum of this location is the luminosity-weighted sum of these 
components:

\begin{displaymath}
S_{loc} = f \cdot S_{*} + (1-f) \cdot S_{other}
\end{displaymath}

To measure the velocity dispersion the spectrum of the target location is convolved with that 
of a reference star, S$_{ref}$:

\begin{displaymath}
S_{loc} \star S_{ref} = (f \cdot S_{*} + (1-f) \cdot S_{other} ) \star S_{ref}
\end{displaymath}

\noindent{where convolution is indicated with the symbol $\star$.} The Fourier transform 
of the cross-correlation function is then:

\begin{displaymath}
F(S_{loc} \star S_{ref}) = \overline{F}(f \cdot S_{*} + (1-f) \cdot S_{other}) \cdot F(S_{ref})
\end{displaymath}

\noindent where the cross-correlation theorum has been applied, and $\overline{F}$ denotes 
complex conjugation. This reduces to:

\begin{displaymath}
F(S_{loc} \star S_{ref}) = f \cdot \overline{F}(S_{*}) \cdot F(S_{ref}) + (1-f) \cdot \overline{F}(S_{other}) \cdot F(S_{ref})
\end{displaymath}

Applying the inverse transform then:

\begin{displaymath}
S_{loc} \star S_{ref} = f \cdot (S_{*} \star S_{ref}) + (1-f) \cdot (S_{other} \star S_{ref})
\end{displaymath}

The cross correlation function of the composite spectrum with 
the reference star is thus the sum of the cross correlation 
functions of the components of the target spectrum with the reference star. The properties of 
the final cross-correlation function then depend on the properties of the various sources 
that contribute to the total light, their degree of similarity, 
not only among themselves but with the reference star, and 
the fractional contribution that each component makes to the total light. 

	In principle, a threshold value of the measured velocity 
dispersion that corresponds to some pre-defined value of $f$ can be determined from 
simulations, in which a spectrum corresponding to that of the dominant light 
source are added in various proportions to a spectrum representing 
the galaxy body that has been smeared to simulate 
the impact of velocity dispersion. The situation is simplified somewhat in 
near-infrared studies of intermediate age or older systems, as a large fraction of 
the $K-$band light comes from the brightest evolved stars. A consequence is that the spectra 
of the brightest resolved stars are similar to that of the underlying integrated light, 
making it is easier to match the reference star spectrum with those of the integrated galaxy 
light and any resolved stars. 

	A basic assumption is that the resolution elements that 
are dominated by individual stars have an underlying stellar component with a 
$\sigma$ that surpasses that which can be measured from the data. Higher spectral 
resolutions will be required to detect single sources in fields with lower 
$\sigma$. It will thus prove more challenging to identify 
locations that are dominated by light from a single source in 
crowded enviroments where the stars have modest (at least when compared with the centers of 
galaxy bulges) random motions, such as globular clusters.

\section{AN APPLICATION TO M32 NIFS SPECTRA}

\subsection{The Identification of Individual Stars}

	The velocity dispersions of the sources are listed in 
the last column of Table 1. The M0III star HR 4884, which was also observed with NIFS, was 
used as the reference template for these measurements. $\sigma$ was measured 
in the $2.1 - 2.3\mu$m interval, which is where the S/N ratio is highest.
The uncertainties in the $\sigma$ measurements are $\leq \pm 10$ km sec$^{-1}$. 
The spectra of sources 10 and 14 were too noisey to allow velocity dispersions to be 
measured.

	There is a broad range in velocity dispersions in Table 1, and half of the objects 
have $\sigma > 90$ km sec$^{-1}$. Given that the velocity 
dispersion of M32 within the central 1.5 arcsec is $\geq 100$ km sec$^{-1}$, then 
the majority of locations in the M32 NIFS spectra that contain 
a distinct point source are actually blends where the light originates from a mix of 
stars, with no single star dominating the light output. The appearance of a point source 
at these locations is simply the result of statistical flucuations in the stellar content. 

	For this study, we identify locations where at least 60\% (i.e. 
$f \geq 0.6$) of the light comes from a single star. Simulations were run in which 
an M5III spectrum (see below) was added in various proportions 
to M giant spectra that were smeared to $\sigma = 100$ km sec$^{-1}$. An M giant 
spectrum was selected to represent the underlying light as this is consistent with 
the integrated $K-$band spectrum of M32 (Davidge 1990). These simulations indicate 
that locations with photometrically-distinct sources that have $\sigma < 35$ km sec$^{-1}$ are 
those where at least 60\% of the light comes from a single star. 

	Three of the sources in Table 1 have $\sigma \leq 35$ km sec$^{-1}$. A fourth 
source is also of interest, as the cross-correlation function with the reference star is 
double-peaked, suggesting that it is dominated by two stars that contribute equally to the light.
The spectra of the sources with $\sigma < 35$ km sec$^{-1}$ and of the two-star blend are 
shown in Figure 2. The spectra have low S/N ratios due to 
noise introduced by the bright unresolved body of fainter stars in M32. 
Still, the S/N ratio is such that the $^{12}$CO band head can be identified. As by far 
the most prominent feature in each spectrum, the $^{12}$CO band head 
dominates the velocity dispersion measurements. The depth of the $^{12}$CO 
band head is a measure of velocity dispersion, and this is demonstrated in Figure 2, where 
the mean spectrum of M32 at 1 arcsec radius from 
Davidge et al. (2008) is shown. The $^{12}$CO band heads in the mean M32 and Star 13 spectra 
have similar depths, as expected given their comparable velocity dispersions. However, the 
$^{12}$CO band heads in the spectra of Stars 1, 3, and 8 are much deeper, reflecting their lower 
velocity dispersions.

	The velocity dispersions of Stars 1, 3, and 8 fall well below the range of 
velocity dispersions that are measured in integrated light throughout the 
central regions of M32. This is demonstrated in Figure 3, which shows the 
histogram distribution of velocity disperions measured in $0.12 \times 0.1$ arcsec$^2$ 
sections within the $3 \times 3$ arcsec$^2$ NIFS field of view. The $0.12 \times 0.1$ 
arcsec$^2$ area corresponds to that subtended by point sources in these data. 

	The histogram in Figure 3 contains $\sigma$ measurements from locations that span a 
range of radii and position angles, and so the distribution is broader than 
expected from random measurement errors alone. In addition, the velocity dispersions 
measured for Stars 1, 3, and 8 are much smaller than what is 
measured in integrated light throughout the central regions of M32. 
Hence, the small velocity dispersions measured for these sources are not 
due to random flucuations in the velocity field.

\subsection{Spectral Types and Line Strengths}

	The low S/N ratio of the extracted spectra notwithstanding, some 
information can be extracted by multiplexing the signal over a range of 
wavelengths, rather than relying on measurements made within narrow wavelength intervals, such 
as line index measurements. A basic piece of information that can be extracted are the spectral 
types of the resolved stars. This is relevant for stellar content studies because 
spectral-type can serve as a proxy for color, allowing comparisons to be made 
with photometric observations at larger radii.

	Spectral types were determined by cross-correlating the M32 spectra with those of 
reference stars from Rayner, Cushing, \& Vacca (2009), which were downloaded from 
the IRTF website. The IRTF spectra were re-sampled and smoothed to match the 
wavelength sampling and spectral resolution of the processed NIFS spectra. The height of 
the central peak in the cross-correlation function measures the degree of similarity 
between the two spectra. The amplitude of the cross-correlation peak can be biased by 
a single strong feature, such as the $^{12}$CO (2,0) band head. Therefore, in 
addition to correlations in the $2.1 - 2.3\mu$m wavelength interval, a second set of 
correlations was done in the $2.1 - 2.28\mu$m interval to avoid the $^{12}$CO band head 
and thereby base the spectral matching on the strengths 
of atomic and less prominent molecular features. Significantly, 
the cross-correlations involving the wavelength intervals with and without the 
$^{12}$CO band head produced consistent results, indicating that the spectral typing is not 
biased by a single strong feature.

	The highest degree of semblance resulted when the spectra in Figure 2 
were compared with the M5III star. The next best fit was for M6III, while 
the peaks of the cross-correlation functions involving types M9III and C2.2 
were substantially lower than those involving M0 III - M6 III stars. The degree of 
similarity between the near-infrared spectra of the M32 sources and the Galactic M5 III 
standard star is demonstrated in Figure 4, where the mean spectrum of sources 1, 3, and 8 
is compared with that of selected stars from Rayner et al. (2009). The $^{12}$CO (2,0) band 
head and the lines of Na I, Fe I, and Ca I in the mean M32 spectrum are similar in strength 
to those in the M5 III spectrum.

	The similarity in spectral-types is significant, as it indicates that the 
upper regions of the AGB in M32 are populated by stars with a narrow range in effective 
temperature. Adopting the relation between spectral type and $J-K$ from Bessell 
\& Brett (1988) for solar neighborhood giants, then these 
sources have $J-K = 1.2 - 1.3$, which is consistent with the color of the upper AGB 
at larger radii in M32 (Davidge \& Jensen 2007).

	The amplitude of the cross-correlation function also contains information about 
the strengths of absorption features, albeit multiplexed over a range of elements. Davidge 
et al. (2008) found that the integrated spectrum of M32 falls near relations defined by 
solar neighborhood stars on the (Ca I, $^{12}$CO) and (Na I, $^{12}$CO) diagrams, suggesting 
that stars in M32 formed from material that experienced slow enrichment; hence, the 
use of solar neighborhood reference stars for abundance analysis is appropriate. 
The dominant features in the $2.1 - 2.28\mu$m wavelength interval are 
Ca, Na, and Fe lines, although other elements also contribute to these features (Silva, 
Kuntschner, \& Lyubenova 2008). When cross-correlated with an M5III star in this 
wavelength range, the peaks measured for stars 3, 8, and 13 are very similar. However, the 
peak of the Star 1 cross-correlation function is $\sim 0.6 \times$ those of the other 
stars, suggesting that Star 1 has weaker atomic absorption 
lines than the other sources. It should be emphasized that this measurement is not 
based on any one feature in the spectrum, but the depth of all of the features in the 
$2.1 - 2.28\mu$m interval, and hence is best understood as an overall metallity 
difference, such that the mean metallicity in Star 1 is $\approx 0.2$ dex lower than the 
other stars. This is well within the dispersion that is seen in other spheroidal systems, 
such as the Galactic Bulge (e.g. Zoccali et al. 2003), and falls within the metallicity 
dispersion that has been proposed to explain the width of the AGB at larger radii in M32 
(Davidge \& Jensen 2007).

\section{SUMMARY \& DISCUSSION}

	We have discussed the use of absorption line widths to identify locations in crowded 
fields where the light is dominated by a single object. The use of line widths to identify 
such objects is most effective in kinematically hot environments, where the line width 
contrast between obvious blends and locations where the light is dominated by a single star is 
greatest. We have examined $2.1 - 2.3\mu$m spectra with an angular 
resolution of $\approx 0.1$ arcsec in the central few arcsec of the cE galaxy M32. 
An initial list of potential individual stars was defined by identifying 
photometrically distinct sources. The majority of these are found to have 
line widths that are consistent with the light originating from a mix of stars with 
a velocity dispersion that is comparable to that of the main body of 
the galaxy; the extracted spectra are thus of composite 
systems made up of physically unrelated stars that, due to stochastic effects, 
appear as compact sources at this angular resolution.
As for the four remaining sources, three have line widths indicating 
that the light is dominated by a single intrinsically luminous source, while the fourth 
has a double-peaked cross-correlation profile that is indicative of two objects with 
similar $K-$band magnitudes that are isolated in velocity space. 

	There are hints from other datasets 
that point sources with light coming largely from one star should be found in these data. 
Davidge et al. (2000) used images with an angular resolution of 0.12 arcsec to 
investigate the brightest stars near the center of M32. While their 
$(K, H-K)$ CMD of the 0 -- 3 arcsec interval has a conspicuous spray of bright sources that are 
obvious blends, a weak concentration of AGB-tip stars can be seen at $K \approx 15.5$, 
which is the AGB-tip magnitude seen at larger radii. These are point sources close to the 
center of M32 where the light is probably dominated by a single bright AGB-tip star.

	It can be anticipated that, when compared with blends, unblended sources 
(1) will be fainter, and (2) will be in areas of lower projected stellar 
density. The brightnesses and local background levels of all fourteen 
locations were measured from the $K$ images discussed by Davidge et al. (2000). 
The locations with broad lines tend to be almost $2 \times$ brighter in $K$ 
than those with narrow lines, and are in regions where the
background within 0.2 arcsec of the source is $\sim 10\%$ higher. With the 
caveat that a large fraction of the brightest stars are LPVs (Davidge 
\& Rigaut 2004), the sources with low velocity dispersions thus tend to be fainter than the 
high velocity-dispersion sources and are located in areas with lower background levels, 
as expected if they are not blends of many objects with comparable brightness.

	While the sample of single stars detected here is small, modest insights 
can be gained into the bright stellar content near the center of M32. The objects that 
are dominated by one or two bright stars are almost certainly evolving near the 
AGB-tip, and these all have an M5III spectral type. These spectral type measurements are 
not biased by a single strong feature, such as the (2,0) $^{12}$CO band head. 
The uniformity in spectral type is consistent with the distinct red cut-off in the 
$(K, H-K)$ CMD of the central regions of M32 found by Davidge et al. (2000). It also 
suggests that the $J-K$ color of the AGB-tip near the center of M32 is consistent with 
that at large radii, indicating that the homogeneous nature of the AGB sequence in 
M32 measured by Davidge \& Jensen (2007) extends into the central arcsec of the galaxy.

	The spectra also suggest that there are modest star-to-star metallicity differences 
among the brightest stars in M32. When the narrow-line spectra are cross-correlated with a 
reference spectrum in the $2.1 - 2.28\mu$m interval, one source -- \# 1 -- has a significantly 
lower peak in the correlation function, by an amount that suggests it has a metallicity that is 
$\approx 0.2$ dex lower than the other stars. Metallicity differences of this size fall well 
within the spread seen in other spheroidal systems, and are consistent with the spread in $J-K$ 
color seen amongst AGB stars in the outer region of M32 (Davidge \& Jensen 2007).

	Many of the bright AGB stars in M32 are LPVs (Davidge \& Rigaut 2004), and there is 
an obvious bias to detect these objects when they are brightest. 
Observations at future epochs may find new locations near the center of M32 that are 
dominated by single LPVs that are near the peak of their light variation, but were in a part of 
their light curve that rendered them too faint to be resolved in the present data. By the 
same token, the sources found here may not be detected if they are variable and in a faint 
phase of their light cycle at the time of future observation. This raises the possibility that 
the sample of spectroscopically-identified stars near the center of M32 can be increased by 
observing at different epochs. This will provide a larger sample of objects from 
which to measure, for example, a more robust metallicity dispersion.

\acknowledgements{Thanks are extended to an anonymous referee for suggesting the comparison 
shown in Figure 3.}

\clearpage

\begin{table*}
\begin{center}
\begin{tabular}{ccl}
\tableline\tableline
\# & R$_{Nuc}$ & $\sigma$ \\
 & (arcsec) & (km sec$^{-1}$) \\
\tableline
1 & 1.34 & 0 \\
2 & 1.46 & 96 \\
3 & 1.58 & 35 \\
4 & 1.45 & 65 \\
5 & 1.10 & 120 \\
6 & 0.96 & 161 \\
7 & 0.87 & 62 \\
8 & 1.33 & 31 \\
9 & 1.14 & 106 \\
10 & 0.95 & -- \\
11 & 1.15 & 114 \\
12 & 1.27 & 80 \\
13 & 1.52 & 96 (Bimodal)\\
14 & 1.53 & -- \\
\tableline
\end{tabular}
\end{center}
\caption{Galactocentric Radii and Velocity Dispersion of the Sources Studied}
\end{table*}

\clearpage

\begin{figure}
\figurenum{1}
\epsscale{0.75}
\plotone{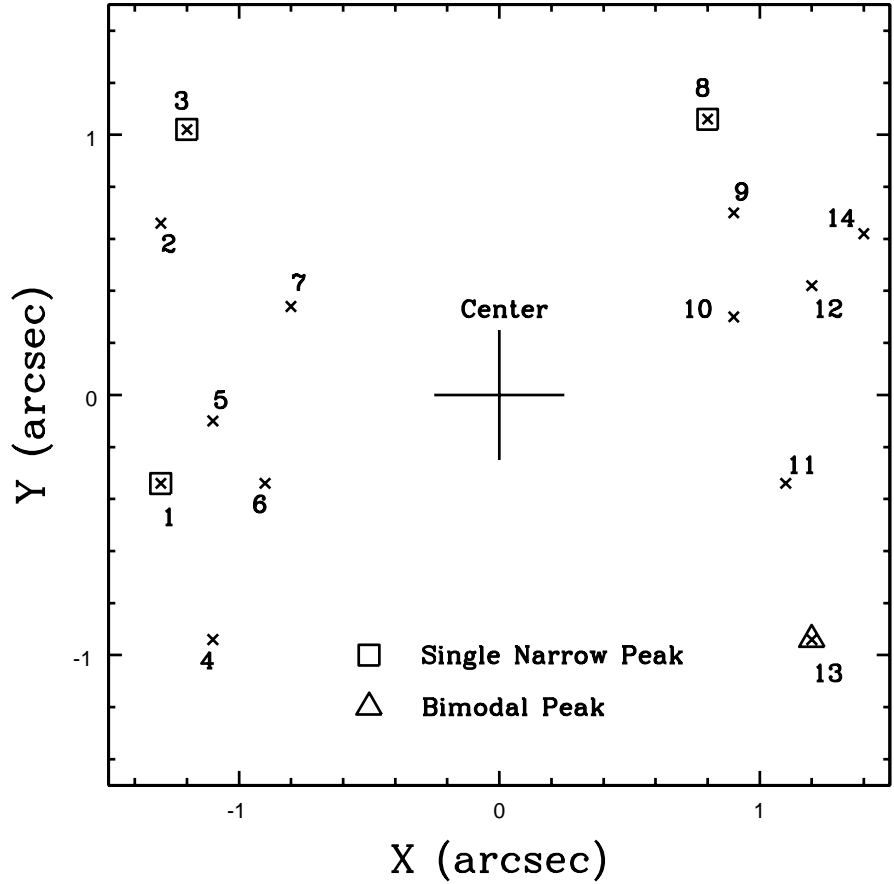}
\caption{The locations of point sources for which spectra were extracted. $X$ and $Y$ are 
offsets from the center of M32, with North at the top and East to the left. 
Sources that have a velocity dispersion $\leq 35$ km sec$^{-1}$, indicating 
that a single bright object contributes at least 60\% of the light, are marked with a box. A 
source with a two-peaked cross-correlation function, suggesting that the light is 
dominated by two stars of comparable brightness, is indicated with the triangle.}
\end{figure}

\clearpage

\begin{figure}
\figurenum{2}
\epsscale{0.75}
\plotone{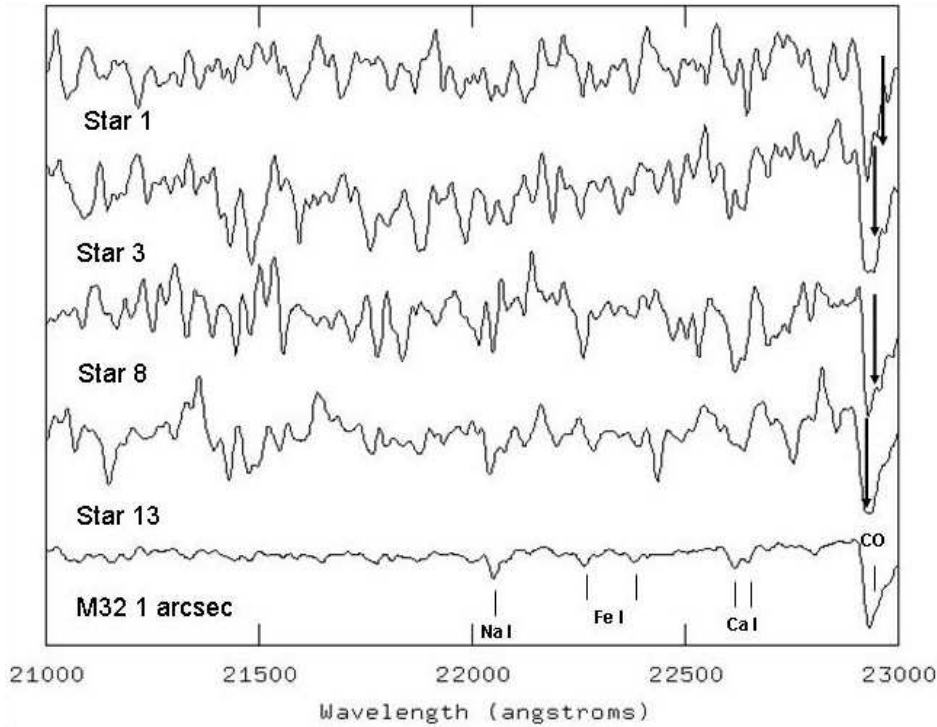}
\caption{Spectra of the sources near the center of M32 in which the $K-$band light is 
dominated by one (Sources 1, 3, and 8) or two (Source 13) bright stars. 
The mean spectrum of M32 at 1 arcsec radius from Davidge et al. (2008) is also shown. 
The locations of various atomic features, and the first-overtone (2,0) 
$^{12}$CO band head are indicated. Note that the Ca I and Na I lines in Star 1 appear to be 
weaker than in the other 3 sources. The arrows show the depth of the $^{12}$CO band head in 
the composite M32 spectrum. The $^{12}$CO band head in the Star 13 spectrum is almost as deep 
as in the M32 spectrum, while the $^{12}$CO band head in the Star 1, 3, and 8 spectra are much 
deeper. These differences in $^{12}$CO band head depth are due to differences 
in velocity dispersion.}
\end{figure}

\clearpage

\begin{figure}
\figurenum{3}
\epsscale{0.75}
\plotone{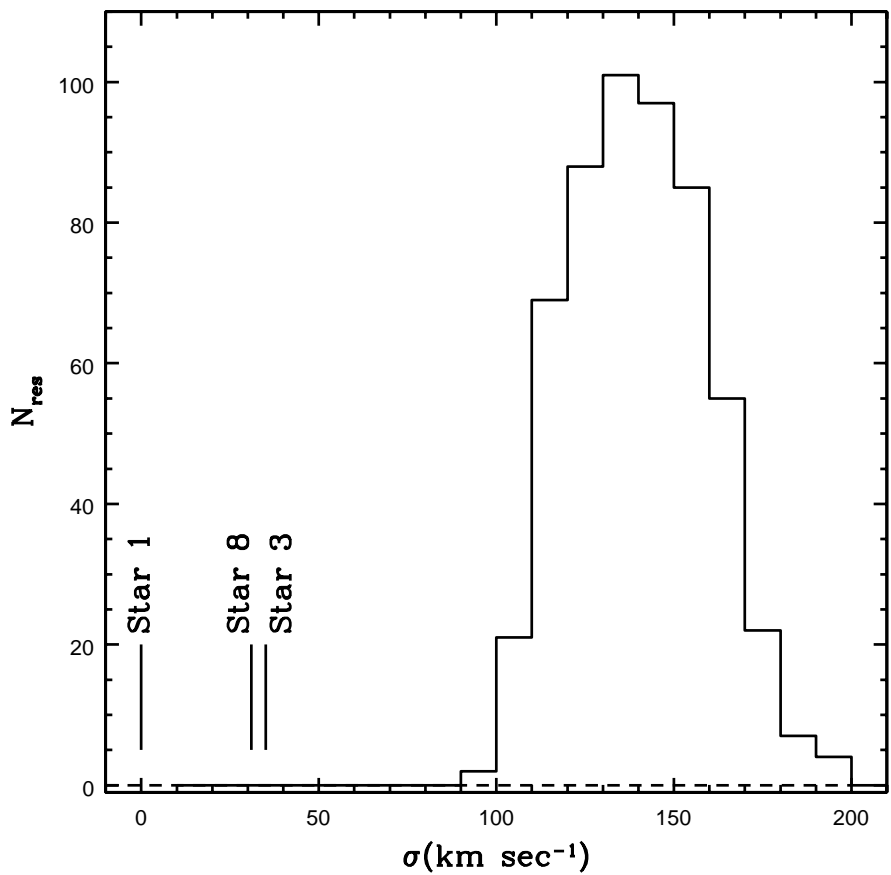}
\caption{The histogram distribution of velocity dispersion measurements of integrated light 
throughout the $3 \times 3$ arcsec$^2$ NIFS field of view. Velocity dispersions were measured 
in $0.12 \times 0.1$ arcsec$^2$ regions, which is the approximate area subtended by 
point sources in these data. $N_{res}$ is the number of angular resolution 
elements that have a velocity dispersion within each 10 
km sec$^{-1}$ interval. The velocity dispersions measured for Stars 1, 
3, and 8 are also indicated. These objects have velocity dispersions that fall well below 
the range of velocity dispersions measured in the NIFS field of view, indicating that 
their low velocity dispersions are not a consequence of statistical flucuations in the 
velocity field.}
\end{figure}

\clearpage

\begin{figure}
\figurenum{4}
\epsscale{0.75}
\plotone{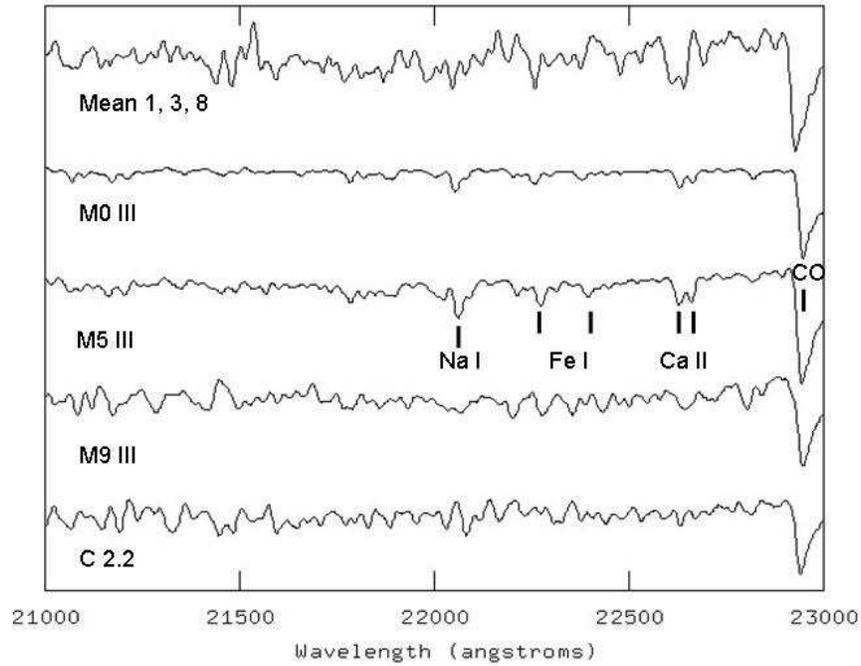}
\caption{The mean spectrum of the three stars with $\sigma \leq 35$ km sec$^{-1}$ 
is compared with spectra of various Galactic stars. A cross-correlation analysis 
indicates that the spectra of the M32 stars have the greatest 
similarity with the M5 III spectrum. Note that the 
atomic features that dominate the wavelength region shortward of $2.28\mu$m in early 
to mid-M giants largely disappear in the M9III and C star spectra.} 
\end{figure}

\end{document}